The Creation and Propagation of Radiation: Fields Inside and Outside of Sources

Stanislaw Olbert and John W. Belcher
Department of Physics
Massachusetts Institute of Technology

Richard H. Price
Department of Physics
University of Texas at Brownsville

**Abstract**

We present a new algorithm for computing the electromagnetic fields of currents inside and outside of finite current sources, for arbitrary time variations in the currents. Unexpectedly, we find that our solutions for these fields are free of the concepts of differential calculus, in that our solutions only involve the currents and their time integrals, and do not involve the time derivatives of the currents. As examples, we give the solutions for two configurations of current: a planar solenoid and a rotating spherical shell carrying a uniform charge density. For slow time variations in the currents, we show that our general solutions reduce to the standard expressions for the fields in classic magnetic dipole radiation. In the limit of extremely fast turn-on of the currents, we show that for our general solutions the amount of energy radiated is exactly equal to the magnetic energy stored in the static fields a long time after current creation. We give three associated problem statements which can be used in courses at the undergraduate level, and one problem statement suitable for courses at the graduate level. These problems are of physical interest because: (1) they show that current systems of finite extent can radiate even during time intervals when the currents are constant; (2) they explicitly display transit time delays across a source associated with its finite dimensions; and (3) they allow students to see directly the origin of the reaction forces for time-varying systems.

# 1    Introduction

There are almost no analytic solutions for the complete electromagnetic fields everywhere in space due to time-varying current systems with at least one finite and non-zero spatial dimension.  In principle, such problems can be solved numerically, with the standard approach being first to solve the time-harmonic problem, and then to find the solution for arbitrary time-dependence using inverse Fourier transforms [1].  However, in the general case, this approach involves numerical computations of substantial complexity.  As a result, we avoid in the intermediate level classroom discussions of the fields associated with such problems.  And even in graduate courses, where such problems may be considered, the physical meaning of the solutions are usually lost in a maze of mathematical complexity (e.g., the solutions to the general problem involve vector spherical harmonics).

However, we would like a student in intermediate and graduate level courses to be able to solve analytically problems where an abrupt change in current in one part of a current system causes changes in the local fields, and where these field changes can then be seen to propagate at the speed of light to other parts of the same system of currents, and interact with those currents, as well as eventually producing radiation far away from the source.  Among many pedagogical advantages, such problems directly illuminate the nature of reaction forces from first principles, instead of deriving the form for those forces from energy considerations.  They also show that finite current systems can radiate even during time intervals where the currents are not changing in time, a result that cannot be understood for a point source, e.g. classic magnetic dipole radiation from a point dipole.

We are thus motivated to look at the properties of the complete analytic solutions to two systems of the type described above:  a planar solenoid with the distance between the infinite planes of current given by $2a$; and a rotating, uniformly-charged spherical shell with radius $a$.  For both of these problems, we give the complete vector potential everywhere in space, for arbitrary time dependence of the currents.  When we go to the slow-motion or dipole approximation limit, our complete solutions illuminate properties of magnetic dipole radiation that are not accessible in the classic magnetic dipole radiation solutions.  Our complete solutions also yield new information which was previously inaccessible about the energy radiated for "instantaneous" turn-on.  We give three problem statements for the undergraduate level related to the properties of these solutions.  We give a fourth problem statement, for a related spherical shell of current problem, appropriate for a graduate level course in electromagnetism.

We note that the kinds of models we consider here, especially that of the spherical shell, were at the forefront of scientific research one hundred years ago in the Abraham-Lorentz models of the electron.   Far from being a moribund field of study, the Abraham-Lorentz models of the spinning electron have recently been rediscovered as a fascinating dynamical system [2].  However, our purpose here is much more prosaic than current research addressing these dynamics.  Here we assume the motions are given, and we focus on causality, reaction forces, and transit time effects in the consequent generation of electromagnetic fields.

The remainder of this paper is organized as follows: The planar solution is derived in Sec. 2, and the solution for the rotating spherical shell is sketched in Sec. 3. These solutions are used in Sec. 4 to examine issues of propagation, radiation without acceleration, what is hidden by the slow-motion or dipole approximation, energy considerations, local reaction forces, and more. Sec. 5 gives a summary of results and insights, and Sec. 6 presents problems suitable for courses in electromagnetism.

## 2    The planar solenoid

The problem of the fields of an infinite parallel current sheet changing arbitrarily in time has been treated by many authors [3]. We give a brief derivation of those fields here, using an approach which can also be applied to the problem of the rotating spherical shell. The differential equation for the vector potential $\mathbf{A}$ in the Lorentz gauge is

$$-\nabla \times \nabla \times \mathbf{A} - \frac{1}{c^2}\frac{\partial^2 \mathbf{A}}{\partial t^2} = -\mu_o \mathbf{J} \tag{1}$$

In the case of a current sheet in the $y$-direction located at $x = 0$, our current density is given by

$$\mathbf{J}(x,t) = \kappa(t)\delta(x)\hat{\mathbf{y}} \tag{2}$$

where $\kappa(t)$ is the current per unit length and is an arbitrary function of time. For this current density, the vector potential $\mathbf{A}$ has only a $y$-component. With Eq. (2), Eq. (1) becomes

$$\left(\frac{\partial^2}{\partial x^2} - \frac{1}{c^2}\frac{\partial^2}{\partial t^2}\right)A_y(x,t) = -\mu_o \kappa(t)\delta(x) \tag{3}$$

The general solution to the homogeneous equation in Eq.(3) is of the form

$$G(t - x/c) + H(t + x/c) \tag{4}$$

where $G$ and $H$ are any functions of a single variable. To find the particular solution to Eq.(3) with the inhomogeneous source term, we build that solution out of solutions to the homogeneous equation. Using causality, we assume a function propagating to the left for $x < 0$ and to the right for $x > 0$. Continuity at the origin then requires that

$$A_y(x,t) = G\left(t - |x|/c\right) \tag{5}$$

To determine $G(t)$, we note that Eq. (3) implies that we must have a jump in the $z$-component of the magnetic field (or in the $x$ derivative of $A_y$) across $x = 0$ given by

$$B_z \big|_{x=+\varepsilon} - \quad B_z \big|_{x=-\varepsilon} = \frac{\partial A_y}{\partial x}\bigg|_{x=+\varepsilon} - \frac{\partial A_y}{\partial x}\bigg|_{x=-\varepsilon} = -\mu_o \kappa(t) = -\frac{2}{c}\frac{dG(t)}{dt} \tag{6}$$

We have derived the last term on the right of Eq.(6) using Eq.(5) to evaluate the left-most expression in Eq. (6). Thus we have

$$A_y(x,t) = \frac{c\mu_o}{2}\kappa^{(1)}\left(t - \frac{|x|}{c}\right) + C \tag{7}$$

where

$$\kappa^{(1)}(t) = \int_0^t \kappa(t')\,dt' \tag{8}$$

We note that our solution for the vector potential depends on the *first time integral* of the current as a function of time, and not on any of its time derivatives. We will see a similar behavior in the rotating spherical shell case, except we will have terms up to the *third* time integral of the current as a function of time in that case (cf. Eq. (19) below).

The electric and magnetic fields, given by the spatial and temporal derivatives of the vector potential, will thus be proportional to $\kappa(t)$ and not to any of its time derivatives [4]. Thus the current sheet can radiate energy to infinity even when the current per unit length $\kappa(t)$ is constant in time. One might suspect that this strange behavior (radiation without acceleration) is somehow a feature of the infinite planar geometry, and would not appear in more realistic geometries. However, we will obtain the same behavior in the case of a rotating spherical shell of charge, as we discuss below in Sec. 4. In that discussion we explain why this behavior (radiation without acceleration) is in fact understandable physically in both cases.

We now use superposition to write down the solution for a planar solenoid. By "planar solenoid" we mean that we have two current sheets in the $yz$ plane, one located at $x = a$ and one located at $x = -a$, with the current sheet at $x = a$ given by $\hat{\mathbf{y}}\,\kappa(t)$ and the current sheet at $x = -a$ given by $-\hat{\mathbf{y}}\,\kappa(t)$. The solution for this case is a superposition of the two solutions of the type given in Eq. (7) and can be written in the form

$$\mathbf{A}(x,t) = \hat{\mathbf{y}}\,\frac{c\mu_o}{2}\left\{\kappa^{(1)}\left(t - \frac{|x-a|}{c}\right) - \kappa^{(1)}\left(t - \frac{|x+a|}{c}\right)\right\} \tag{9}$$

## 3    The rotating spherical shell of charge

We turn to the problem of a uniformly-charged rotating shell of radius $a$. We assume that the radius of the sphere and the distribution of charge remain constant even as the sphere rotates. This is impossible, of course, since there are no materials that are perfectly stiff and perfectly insulating, but this is not relevant here since our focus is on causality in the generation of electromagnetic fields. Our rotating shell carries current in the azimuthal direction, with the current depending on the sine of the polar angle $\theta$. Solutions to this problem in a very different form from ours have been given by Daboul and Jensen, and also by Vlasov [5]. Our current density is given by

$$\mathbf{J}(r,\theta,t) = \hat{\boldsymbol{\phi}}\,\kappa(t)\,\delta(r-a)\sin\theta \tag{10}$$

If the sphere has charge per unit area $\sigma$ and rotates with an angular speed $\Omega(t)$, then $\kappa(t) = \sigma\,a\,\Omega(t)$. For future use, we note that the magnetic dipole moment of the current distribution given in Eq. (10) is

$$\mathbf{m}(t) = \frac{1}{2}\int \mathbf{r}' \times \mathbf{J}(\mathbf{r}',t)\,d^3x' = \hat{\mathbf{z}}\,\frac{4\pi a^3}{3}\kappa(t) \tag{11}$$

If the sphere has a total charge $Q$ and is rotating at a fixed angular rotation rate $\Omega_o$, then we have that the static magnetic dipole moment $m_o$ is given by

$$m_o = \frac{4\pi}{3}a^3\kappa_o = \frac{a^2\Omega_0 Q}{3} \tag{12}$$

The vector potential $\mathbf{A}$ for the current density given in Eq. (10) only has a $\phi$-component, and $A_\phi$ satisfies the differential equation

$$\frac{1}{r}\frac{\partial^2}{\partial r^2}\left(rA_\phi\right) + \frac{1}{r^2}\frac{\partial}{\partial\theta}\left[\frac{1}{\sin\theta}\frac{\partial}{\partial\theta}\left(\sin\theta A_\phi\right)\right] - \frac{1}{c^2}\frac{\partial^2}{\partial t^2}A_\phi = -\mu_o J_\phi \tag{13}$$

For the $\sin\theta$ dependence of $\mathbf{J}$ in Eq. (10), $A_\phi(r,\theta,t)$ can be written in separable form as

$$A_\phi(r,\theta,t) = A(r,t)\sin\theta \tag{14}$$

We find the differential equation for $A(r,t)$ by inserting Eq. (14) into Eq. (13) and using Eq. (10), yielding

$$\frac{\partial^2}{\partial r^2}(rA) - \frac{2}{r^2}(rA) - \frac{1}{c^2}\frac{\partial^2}{\partial t^2}(rA) = -\delta(r-a)\mu_o\kappa(t) \tag{15}$$

The reader should compare Eq. (15) for the spherical case to Eq. (3) for the planar case. The general solution to the homogeneous equation in Eq.(15) is of the form

$$r\,A(r,t) = G'(t - r/c) + \frac{cG(t - r/c)}{r} + H'(t + r/c) - \frac{cH(t + r/c)}{r} \tag{16}$$

As in the planar case, we find the particular solution to Eq.(15) by assuming the form in Eq. (16), but now we use causality to assume propagation both inward and outward for $r < a$ and only outward for $r > a$, and we require regularity of the solution at $r = 0$. We then require continuity of $A(r,t)$ at $r = a$ as well as the jump in the $r$-derivative of $A(r,t)$ at $r = a$ implied by the delta function in the source term in Eq. (15). As in the planar case, these requirements suffice to completely determine our particular solution.

We can also find the solution to Eq. (15) by Laplace transforming and inverting. Since the details of either process are involved, and we are mainly interested in the properties of the solutions in any case, we relegate the details of both the time-domain and the Laplace transform approach to a technical note [6], and only state the results here. We use Eq. (11) to write our solution in terms of $m(t)$ instead of $\kappa(t)$, to facilitate comparison with the classic magnetic dipole solutions, which involve $m(t)$ and its time derivatives. If we define

$$r_< = \min(r, a) \qquad r_> = \max(r, a) \tag{17}$$

and

$$t_\pm = t - \frac{r_>}{c} \pm \frac{r_<}{c} \tag{18}$$

then we can write the solution for the vector potential of our spherical shell, which only has a $\phi$ component, as

$$A_\phi^{\text{shell}}(r, \theta, t) = \frac{3c\mu_o \sin\theta}{8\pi a^2 r} \left\{ \begin{array}{l} \left[ m^{(1)}(t_+) + m^{(1)}(t_-) \right] - \dfrac{c}{r_<} \left[ m^{(2)}(t_+) - m^{(2)}(t_-) \right] \\[2mm] + \dfrac{c}{r_>} \left\{ \left[ m^{(2)}(t_+) + m^{(2)}(t_-) \right] - \dfrac{c}{r_<} \left[ m^{(3)}(t_+) - m^{(3)}(t_-) \right] \right\} \end{array} \right\} \tag{19}$$

where we have defined the successive integrals of the dipole moment $m(t)$ by

$$m^{(0)}(t) = m(t) = \frac{4\pi a^3}{3} \kappa(t) \tag{20}$$

$$m^{(n+1)}(t) = \int_0^t m^{(n)}(t')\,dt' \qquad n = 0, 1, 2\ldots \tag{21}$$

Note that even for a constant magnetic dipole moment, the expression for the vector potential in Eq. (19) appears at first glance to be a function of time, because the successive time integrals of a constant $m(t)$ in Eq. (21) are functions of time, even if the magnetic dipole moment itself is not. In fact, this is only a problem at first glance, and a closer examination of Eq. (19) reveals the following. Suppose that the magnetic dipole moment $m(t)$ is zero for $t < 0$, then varies in the time interval $0 \leq t < T$ in an arbitrary manner, and then reaches a constant value $m_o$ for $t \geq T$. Then it can be shown that [6]

$$A_\phi^{\text{shell}}(r,\theta,t) = \begin{cases} 0 & t < \dfrac{|r-a|}{c} \\ \dfrac{\mu_o m_o \sin \theta}{4\pi a^2} \dfrac{r_<^2}{r\, r_>} & t > \dfrac{(r+a)}{c} + T \end{cases} \tag{22}$$

At very long times, this is the correct solution for the vector potential of a uniformly-charged spherical shell spinning at a constant rate, as we expect. We only see time changes in the vector potential at a given radius $r$ in the time interval

$$\frac{|r-a|}{c} \leq t \leq \frac{(r+a)}{c} + T \tag{23}$$

For instantaneous spin-up ($T = 0$), if $r > a$ the observer will only see fields changing in time over a time $2a/c$, and if $r < a$, the observer will only see fields changing in time over a time $2r/c$. For future use, we also define

$$m^{(-n)}(t) = \frac{d^n m}{dt^n} \qquad n = 1, 2 \ldots \tag{24}$$

We will switch back and forth between the notation in Eq. (24) and the notation in which we use dots over the function to denote time differentiation, for example

$$\ddot{m}(t) = \frac{d^2 m(t)}{dt^2} = m^{(-2)}(t) \tag{25}$$

The associated electric and magnetic fields can be found by taking the negative time derivative and the curl of $\mathbf{A}^{\text{shell}}$, respectively. We give the explicit forms for these fields in [6]. We also show in [6] that the solution above in Eq. (19) is identical to the one given by the authors in [5], although upon first examination they look very different.

## 4    Discussion of the complete field solutions

The remarkable thing about our solution for the spherical shell potential in Eq. (19) is that it does not involve any time derivatives of the magnetic dipole moment $m(t)$. Instead, the vector potential and the associated fields are proportional to $m(t)$ itself and its time integrals, evaluated at two different times. In the context of our experience with the classic theory of magnetic dipole radiation, which we review below, this is a totally unexpected result. There are three obvious questions that arise.

(a) Can we recover the classic magnetic dipole radiation formulae from our complete spherical shell solution when we go to the slow-motion or dipole approximation limit, where we assume that the speed of light transit time across the sphere is negligibly small compared to the time scale for variation in the source?

(b) Can our complete solution to the spherical shell case when considered in the slow-motion or dipole approximation limit, illuminate features of magnetic dipole radiation that we cannot answer in the classic slow-motion or dipole approximation solution?

(c) Is the planar solenoid solution we give above contained in the spherical shell solution in some limit? In particular, can we find situations in the spherical shell case, as we do in the planar solenoid case, where the sphere radiates even though it is spinning at a constant rate? If so, what do these "radiation without acceleration" solutions mean physically?

We consider each of these three questions in turn. First, though, we review the standard development leading to classic magnetic dipole radiation. In that development, we begin with the general solution for the vector potential using the free space Green's function. We then assume that $\mathbf{J}$ vanishes outside of a sphere of radius $a$ and that we are far away from that region. We also assume that if $T$ is the time scale for changes in $\mathbf{J}$, then

$$T \gg a/c \tag{26}$$

We call this approximation the slow-motion or dipole approximation. It is also called the long wavelength approximation, since in this approximation the wavelength of the radiation generated is much larger than the dimensions of the source. We assume that $\mathbf{J}$ is well behaved, in that we can use a Taylor series expansion for the time dependence of $\mathbf{J}$ about the time $t - r/c$. If we also assume that the dipole moment $\mathbf{m}(t)$ is always along the $z$-axis, then to first-order in the small quantities $a/r$ and $a/cT$, the vector potential only has a $\phi$ component, which can be shown to be [7]

$$A_\phi^{\text{classic}}(r,\theta,t) = \frac{\mu_o \sin\theta}{4\pi} \left( \frac{m(t-r/c)}{r^2} + \frac{\dot{m}(t-r/c)}{rc} \right) \tag{27}$$

The electric field and magnetic fields of the classic magnetic dipole can be found from Eq. (27) by taking the negative time derivative of and the curl of the vector potential, respectively. We give explicit forms for these fields in [6].

Note that in the classical slow-motion or dipole approximation for the potential, we have shrunk the radius $a$ of the sphere to zero. The only information we have about the spatial structure of the current distribution inside $a$ is its spatial moments, through $\mathbf{m}(t)$, but we cannot "peer" into the structure itself and we have no idea how the fields behave for $r$ less than or on the order of $a$. The corresponding total rate at which energy is radiated to infinity into all solid angles in classic magnetic dipole radiation is given by

$$\lim_{r \to \infty} \int \left[ \frac{\mathbf{E}^{\text{classic}} \times \mathbf{B}^{\text{classic}}}{\mu_o} \right] \cdot \hat{\mathbf{r}} \; r^2 \sin\theta \, d\theta \, d\phi = \frac{\mu_o \ddot{m}^2}{6\pi c^3} \tag{28}$$

Now let us turn to our three questions above, first considering question (a) (can we recover the classic magnetic dipole radiation formula from our complete spherical shell solution?). It is by no means obvious at first glance that we can do this, since the complete spherical shell solution involves $m(t)$ and its first, second and third time *integrals*, whereas the classic magnetic dipole radiation involves $m(t)$ and its time *derivatives*. But let us assume that all of the time derivatives of $m(t)$ are well behaved, and that we can expand $m(t)$ in a Taylor series. If we look at Eq. (19) and expand about the time $t' = t - r_> / c$, then we can show using the standard Taylor series expansion that this equation can be written as

$$A_\phi^{\text{shell}}(r, \theta, t) = \frac{3 c \mu_o \sin\theta}{4\pi a^2 r} \sum_{k=1}^{\infty} \frac{2k}{(2k+1)!} \left( \frac{r_<}{c} \right)^{2k} \left[ \frac{c \, m^{(2-2k)} \left( t - \frac{r_>}{c} \right)}{r_>} + m^{(1-2k)} \left( t - \frac{r_>}{c} \right) \right] \tag{29}$$

Eq. (29) holds both inside and outside the spherical shell. If we assume that $r \geq a$, then we have

$$\left. A_\phi^{\text{shell}}(r, \theta, t) \right|_{r \geq a} = \frac{3 c \mu_o \sin\theta}{4\pi a^2 r} \sum_{k=1}^{\infty} \frac{2k}{(2k+1)!} \left( \frac{a}{c} \right)^{2k} \left[ \frac{c \, m^{(2-2k)} \left( t - \frac{r}{c} \right)}{r} + m^{(1-2k)} \left( t - \frac{r}{c} \right) \right] \tag{30}$$

Keeping only the first two terms in Eq. (30), we have

$$\left. A_\phi^{\text{shell}}(r, \theta, t) \right|_{r \geq a} = \frac{\mu_o \sin\theta}{4\pi} \left\{ \frac{m(t')}{r^2} + \frac{\dot{m}(t')}{rc} + \frac{1}{10} \left( \frac{a}{c} \right)^2 \left[ \frac{\dddot{m}(t')}{r^2} + \frac{\ddddot{m}(t')}{rc} \right] + \ldots \right\} \tag{31}$$

where now $t' = t - r/c$. In the limit that $a$ goes to zero, we see from Eq. (31) that the vector potential for our complete spherical shell solutions reduces to the familiar expression for the potential associated with classic magnetic dipole radiation, as given in Eq. (27).

Now let us turn to question (b) (can our complete solution to the spherical shell case illuminate features of magnetic dipole radiation that we cannot answer in the classic slow-motion or dipole approximation solutions?). The answer to this question is a resounding yes. Most importantly, in our complete solutions, we still retain a finite and non-zero value of $a$ even when the slow-motion or dipole approximation holds. To make clear what this means, if we assume that $a/cT \ll 1$, and that $m(t)$ can be expanded in a Taylor series, then our expression for the electric field evaluated at $r = a$ can be written (setting $r = a$ in Eq.(31) and taking the negative time derivative) as

$$E_\phi^{\text{shell}}(a, \theta, t) = -\frac{\mu_o \sin\theta}{4\pi} \left\{ \frac{\dot{m}(t')}{a^2} + \frac{\ddot{m}(t')}{ac} + \frac{1}{10}\left(\frac{a}{c}\right)^2 \left[ \frac{\dddot{m}(t')}{a^2} + \frac{\ddddot{m}(t')}{ac} \right] + ... \right\} \qquad (32)$$

where now $t' = t - a/c$. If expand $m(t - a/c)$ in a Taylor series about $t$, Eq. (32) becomes

$$E_\phi^{\text{shell}}(a, \theta, t) = -\frac{\mu_o \sin\theta}{4\pi a^2} \left[ \dot{m}(t) - \frac{2}{5}\left(\frac{a}{c}\right)^2 \dddot{m}(t) + \frac{1}{3}\left(\frac{a}{c}\right)^3 \ddddot{m}(t) + ... \right] \qquad (33)$$

Eq. (33) is an expression for the electric field at the surface of the spherical shell that we have no access to in the classic point magnetic dipole radiation solution, since there we have taken the limit that $a$ is zero. A very important consequence of our "non-classical" dipole result in Eq. (33) is related to reaction forces, that is, the forces that resist the spinning up of the shell. In the classical approach these radiation reaction forces must be inferred from energy conservation. But our solution in Eq.(33) describes the fields acting on the spherical shell, and therefore gives us the reaction fields directly. To spin up the sphere, the external agents must provide a force which exactly counterbalances the electric field reaction forces given in Eq. (33).

With Eq. (33), we can calculate explicitly the rate at which work is being done against this reaction force by the external agents spinning up the sphere. This rate is given by the volume integral of $-\mathbf{J}^{\text{shell}} \cdot \mathbf{E}^{\text{shell}}$, where this quantity is zero everywhere except on the surface of the sphere. Explicitly, using Eq. (10) and Eq. (33), we have

$$\frac{dU_{work}}{dt} = \int \left( -\mathbf{J}^{\text{shell}} \cdot \mathbf{E}^{\text{shell}} \right) r^2 d\Omega$$
$$= \frac{\mu_o}{2\pi a^3} \left\{ m(t)\dot{m}(t) - \frac{2}{5}\frac{a^2}{c^2} m(t)\dddot{m}(t) + \frac{a^3}{3c^3} m(t)\ddddot{m}(t) + ... \right\} \qquad (34)$$

Suppose the magnetic dipole moment has been zero for $t < 0$. If the dipole moment then goes from zero to a constant value in a finite time span, and if the first through third derivatives of $m(t)$ are zero at 0 and at $T$, then the first term on the far right in Eq. (34) will yield the energy stored in the static magnetic field (see Eq. (36) below), the second term will integrate to zero, and the third term will give the total energy radiated to infinity (when integrated by parts twice). That is, the total work done by the external agent spinning up the sphere is the sum of the energy carried away to infinity by the radiation and the energy stored in the static field after spin-up.

By directly using the reaction forces, and computing the work done against them, we have discovered something new and interesting: the "reaction" forces are not only those necessary to supply the energy radiated, but also to establish the energy stored in the final static magnetic field of the rotating shell.

In addition to giving us new access to the reaction forces when the slow-motion or dipole approximation holds, our complete spherical shell solutions also illuminate properties of radiation from a sphere when the sphere is spun-up so rapidly that the slow-motion or dipole approximation does not hold. The classic magnetic dipole solutions have nothing to say in this regime, since to derive these solutions it is assumed that our spin-up is slow in the sense that $T \ll a/c$. In contrast, our complete spherical shell solutions do have significant statements to make, since they are derived with no constraint on the spin-up time. To illustrate what this means, we compute the total energy radiated in our complete spherical shell solutions for any value of $a/cT$, and compare it to the amount radiated in the classic magnetic dipole radiation solutions (cf. Eq. (28)). We choose a form for the time dependence of $m(t)$ which has well-behaved time derivatives to all orders, so that we can compute the energy radiated in the classic magnetic dipole radiation expression in Eq. (28). We take

$$m_{smooth}(t) = \frac{m_o}{2}\left[\frac{2}{\pi}\arctan\left(\frac{t}{T/5}\right)+1\right] \qquad (35)$$

To calculate the total energy radiated in the spherical shell solution, we need only keep the inverse radius terms in the electric and magnetic fields associated with $\mathbf{A}^{shell}$ in Eq. (19), and integrate the associated Poynting flux over time and over the surface area of a sphere at infinity. We thereby numerically compute an expression for the total energy radiated by the spherical shell during the smooth spin-up for any value of $a/cT$. We choose to normalize the total energy radiated by the energy contained in our static fields for constant rotation rate for the spherical shell case, which is given by the formula

$$U_{mag} = \frac{\mu_o m_o^2}{4\pi a^3} \qquad (36)$$

The normalized radiated energy for the smooth turn-on for our complete spherical shell solution is plotted as a function of the ratio $a/cT$ in Figure 1, and labeled "Smooth". For comparison, the dashed straight line labeled "Dipole" in Figure 1 is the normalized radiated energy using the usual expression for the magnetic dipole radiation rate in the slow-motion or dipole approximation, given by Eq. (28). If we compute $\ddot{m}$ using the time dependence given in Eq. (35) and integrate the rate at which energy is radiated to infinity in Eq. (28) over all time, we easily see that the classic dipole total energy radiated scales as $\left( a / cT \right)^3$. At low values of the ratio $a / cT$, our spherical shell solution for the total energy radiated, the "Smooth" curve in Figure 1, shows this behavior, and is essentially identical with the classical result for magnetic dipole radiation. However, as the ratio $a / cT$ becomes comparable to and much greater than unity, our numerical result for the energy radiated for the spherical shell solution approaches the constant value $U_{mag}$.

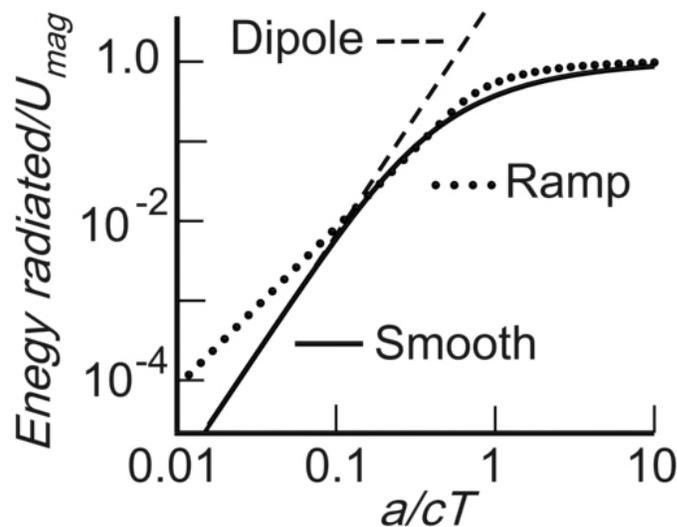

**Figure 1:** The normalized total energy radiated by the spherical shell as a function of $a / cT$ for the smooth spin-up with characteristic time $T$. Ramp spin-up is discussed at the end of Sec. 4. The curve labeled *Dipole* is the total normalized energy radiated by a point magnetic dipole for the smooth spin-up time function.

That is, if we spin up the spherical shell in a time $T$ short compared to $a / c$, we find that the energy radiated away in this process is exactly equal to the energy stored in the magnetic field after the currents reach their terminal values. This is true both for our planar solenoid solutions and for our spherical shell solutions. This also turns out to be true in the spherical shell case not only for current distributions on the sphere proportional to $\sin \theta$, but for any axially-symmetric current distribution with any $\theta$ dependence on the sphere (see Sample Problem 6.4 below).

In contrast, the energy radiated in the classic dipole case as predicted by Eq. (28) increases without limit as $a / cT$ increases. Of course this estimate has no meaning once $a / cT$ becomes comparable to one, since the derivation of the classic equations begins with the assumption that $a / cT \ll 1$. Thus our complete spherical shell solutions illuminate features of the radiation from the spin-up of a spherical shell that are not accessible in the classic magnetic dipole radiation solutions. This is because we can go to the limit of instantaneous turn-on in our complete solutions, whereas the classic magnetic dipole solutions are intrinsically limited to slow turn-on.

Finally, let us turn to question (c) (is the planar solenoid solution contained in some sense in the spherical shell solution?). To answer this question, we must let the radius of the sphere $a$ and the distance from the origin $r$ go to infinity in such a way that the difference $x = r - a$ is finite. In this limit, if we are sitting in the equatorial plane where $\sin\theta = 1$, at a distance $|x| = |r - a| \ll a$ from the spherical surface, the spherical surface looks "planar". To go to this limit we essentially are invoking the reverse of the slow-motion or dipole approximation in Eq. (26). That is we are assuming $T \ll a/c$, rather than the other way around, since we are letting $a$ go to infinity holding $T$ fixed.

We can never approach this limit in the classic magnetic dipole radiation solution, because there we have shrunk the sphere away to a point, and thus we can hardly return to the limit where its radius is finite. But we have no problem taking this limit with our spherical shell solution and the associated fields given in Eq. (19). Consider the equatorial plane at $\theta = \pi/2$, and let $r$ and $a$ go to infinity with $|x| = |r - a| \ll a$ finite. It is straightforward to see that in this limit, the only term to survive in the equation for $E_\phi^{\text{shell}}$ (which is the negative time derivative of the expression in Eq. (19)) is

$$E_\phi^{\text{shell}}(r, \pi/2, t) = -\frac{3c\mu_o}{8\pi a^2 r} m(t - |x|/c) = -\frac{c\mu_o}{2}\kappa(t - |x|/c) \qquad (37)$$

where we have used Eq. (11) to replace $m(t)$ by $4\pi a^3 \kappa(t)/3$, and we have assumed that $m(t)$ is zero in the far distant past. We thus see that in this limit we recover exactly the "strange" behavior of the planar solenoid case, that is, radiation fields which are not zero even when the sphere is rotating at a constant rate.

This result is truly surprising. Even though radiation from a planar sheet moving at constant speed has been discussed previously by a number of authors [3], we are encountering it here for the first time in a more realistic (spherical) geometry. Because we find radiation in this circumstance to be so novel, we revisit the issue of how this can be physically. When the surface is moving at a constant speed, the fields in the vicinity of the surface are not changing in time, yet at and near the surface there is a radiation field $E_\phi$; this is incontrovertibly a radiation field in that this field produces a radiation reaction. Even more surprising is the fact that this locally static radiation reaction field at the shell is associated with changes in the electromagnetic field that are happening at a distance, possibly a large distance, from the shell, the distance to the "wavefront" of the changes that were started with the spin-up process began.

One way of understanding (or at least accepting) that $E_\phi$ is really a radiation field is to return to the more transparent case of the planar sheet of charge and to consider Lorentz transformations for motion in the $y$-direction. If the sheet of charge has never been moving, then in the frame in which the sheet is stationary the electric field has only an $x$ component $E_x$. A Lorentz transformation in the $y$ direction does not change this fact. In a frame in which the sheet is moving in the $y$ direction with some velocity (and has always been moving in the $y$ direction with that same velocity) there is still no $E_y$ component. The $E_y$ component is *not* the result of observing in one frame or another.

This insight can be applied to the case of a sheet that starts from rest in the "lab frame," and then settles down to some constant non-zero speed in that frame. The $E_y$ component cannot be removed by going to a frame co-moving with the sheet at late time. That $E_y$ is irrevocably, irrefutably, and unremovably present. The $E_y$ field, a true radiation field, "knows" not only about the motion of the sheet, but also about its history! These counterintuitive features of the radiation underscore the subtlety of the relationship of radiation and its sources.

To gain additional insight into the physics, let us examine this from another point of view based on momentum transfer for a charged sheet. In the following discussion, we keep only terms to first order in $v_o / c$, but neglect terms of higher order. In the planar case, a single current sheet can be thought of as a moving uniformly-charged sheet with positive charge per unit area $\sigma_o$, moving at velocity $\mathbf{v}(t) = \mathrm{v}(t)\hat{\mathbf{y}}$, with the current per unit length given by $\kappa(t) = \sigma_o \mathrm{v}(t)$. In the discussion of the planar sheet in Sec. 2, we assumed that the moving positive charge $\sigma_o$ is exactly balanced by an stationary negative charge $-\sigma_o$, so that there is no net electrostatic field (see [4] for other possibilities). But consider the case where there is no stationary negative charge, so that we have an electrostatic field. The unbalanced positive charge per unit area $\sigma_o$ produces an electrostatic field of magnitude $\sigma_o / 2\varepsilon_o$ perpendicular to the sheet in the $\pm x$-direction. Suppose that at $t = 0$, we bring this sheet of charge instantaneously from zero speed to constant speed $\mathrm{v}_o$. Then at time $t = T > 0$, for $|x| > cT$ we will simply see the electrostatic field in the $x$-direction, and for $|x| < cT$ we will see both that field and the "radiation" electric and magnetic fields. The ratio of the magnitude of the "radiation" electric field in the $y$-direction to the electrostatic field in the $x$-direction for $|x| < cT$ is given by (cf. Eq. (37))

$$\frac{|E_y|}{|E_x|} = \left[ \frac{c\mu_o \kappa_o}{2} \right] / \left[ \frac{\sigma_o}{2\varepsilon_o} \right] = \frac{\mathrm{v}_o}{c} \qquad (38)$$

This ratio is exactly what we expect from simple geometry. The "foot" of the electrostatic field line is rooted in the charges making up the sheet, and the parts of the field line for $|x| < cT$ are moving along with the "foot" in the $y$-direction at speed $\mathrm{v}_o$ after $t = 0$. Thus we expect from geometry alone to see the ratio given in Eq. (38). This is exactly like waves on a string, and we would have an exactly analogous situation if a post supporting a string is suddenly set in motion with speed $v_o$ perpendicular to the string at $t = 0$, except that the speed $c$ would be replaced by the speed of waves on the string.

   Thus the reason that the unbalanced positive sheet of charge continues to radiate even after it has been brought up to constant speed is that even though the external agents for a time just greater than zero have already put in the energy required to get the sheet itself up to constant speed, they have to continue to do work for $t > 0$ to bring the electrostatic fields associated with the unbalanced charges in the sheet up to speed. The rate at which they continue to add momentum in the $y$-direction per unit area in the $yz$ plane is just the force that they must exert to counterbalance the reaction electric field, that is

$$\left[\sigma_o\right]\left[\left. E_y \right|_{x=0}\right] = \frac{2\varepsilon_o E_o^2}{c^2} v_o c \qquad (39)$$

   Where does the momentum provided by the external agents go? The total electromagnetic momentum in the $y$-direction per unit area in the $yz$ plane at time $t = T > 0$ is

$$2cT\left[\varepsilon_o \mathbf{E} \times \mathbf{B}\right]_y = 2cT\left[\varepsilon_o E_o B_z\right] = 2cT\left[\frac{\varepsilon_o E_o E_y}{c}\right] = T\left[\frac{2\varepsilon_o E_0^2}{c^2} v_o c\right] \qquad (40)$$

Thus we see from Eq. (40) that the electromagnetic momentum in the $y$-direction per unit area in the $yz$ direction is continuously increasing for $t = T > 0$, as the radiation fields move outward at the speed of light, and that the rate at which the electromagnetic momentum is increasing is exactly the rate at which the external agents are providing momentum at $x = 0$ (see Eq. (39)).

   Even long after the unbalanced charged sheet itself is up to speed, the sheet continues to radiate, and that radiation is carrying momentum outward at a rate sufficient to get the more and more distant electrostatic fields "up to speed". For a single sheet this radiation never stops, because there is an infinite amount of electrostatic energy to get up to speed. For our planar solenoid, there is only a finite amount of electrostatic energy per unit area in the $yz$-direction, and those fields are "up to speed" after a time $2a/c$, so that the sheets cease to radiate after that time. We expect exactly this behavior in spinning up our spherical shell of charge.

   From this discussion, it is clear that the situation in which we find radiation without acceleration for the spherical case will occur when the time $T$ in which we spin the sphere up is short compared to the speed of light transit time across the radius of the sphere, and moreover that this "radiation without acceleration" should cease after a time $2a/c$. This is some sense obvious. Suppose we take any finite distribution of current and turn the currents on instantaneously. Far away from this event we would expect to see a burst of radiation that lasts a time of order the finite dimensions of the system divided by the speed of light. Thus the current distribution must be radiating after the currents are no longer changing in time, since the time interval when the currents are changing is zero (or at least arbitrarily small). But even though this is obvious in some sense, we have in Eq. (19) for the first time a complete analytic solution which allows us to examine the process in exact detail, and we explore these details below.

It helps to look at this situation numerically. Consider a value of $a/cT$ which is large compared to 1, but not infinity. For this purpose, we choose a "ramp" turn-on for the time dependence of the magnetic dipole moment, that is

$$m_{ramp}(t) = m_o \begin{cases} 0 & for \ t < 0 \\ \dfrac{t}{T} & for \ 0 < t < T \\ 1 & for \ t > T \end{cases} \tag{41}$$

The total normalized energy radiated during this ramp spin-up function is shown in Figure 1 as a function of $a/cT$ (this is the curve labeled "Ramp"). In Figure 2, we take a value $a/cT = 10$, or $T = 0.1 \ a/c$. We plot $\tilde{E}_\phi^{\text{shell}}$, the $\phi$ component of the electric field of the shell divided by $c\mu_o \kappa_o$, and $\tilde{B}_\theta^{\text{shell}}$, the $\theta$ component of the magnetic field of shell divided by $\mu_o \kappa_o$, at a time $t = 0.5 \ a/c$ after this "ramp" turn-on. Both of the field components are evaluated in the equatorial plane at $\theta = \pi/2$. Figure 2 is one frame of a complete movie that can be found online [6].

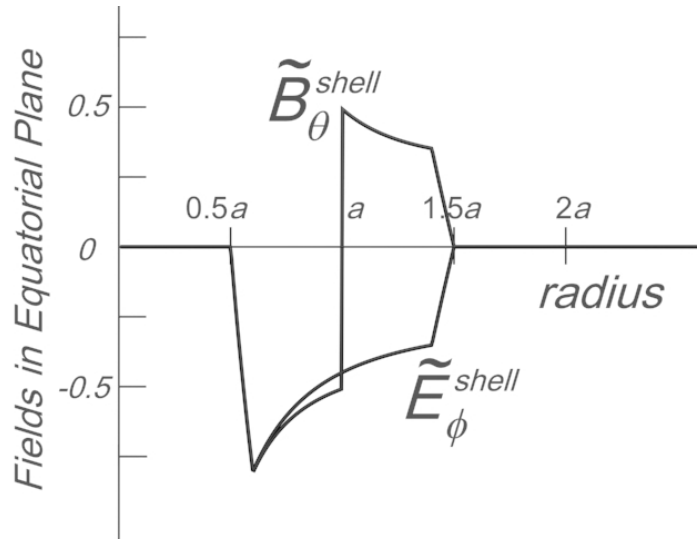

**Figure 2:** The normalized $\phi$ and $\theta$ components of the electric and magnetic field of the shell in the equatorial plane for a ramp turn-on with $T = 0.1 \ a/c$. We show the fields at a time $t = 0.5 \ a/c$ after a turn-on starting at $t = 0$.

What we see in Figure 2 are approximately the fields we would see around a single infinite plane of current located at $x = a$ with a turn-on of the current per unit length given by ramp turn-on of Eq. (41). In the planar case, at $t = 0.5\,a/c$, the normalized electric field would be constant at a value of $-\frac{1}{2}$ in the interval $0.6\,a < x < 1.4\,a$, linearly decreasing from that value to zero at $x = 0.5\,a$ and $x = 1.5\,a$. Similarly, in the planar case at $t = 0.5\,a/c$, the magnitude of the normalized magnetic field would be constant at a value of $\frac{1}{2}$ in the interval $0.6\,a < x < 1.4\,a$, reversing sign across the current sheet at $x = a$, and then linearly decreasing to zero at $x = 0.5\,a$ and $x = 1.5\,a$, respectively. The departures from this planar behavior in Figure 1 are caused by the spherical geometry, but the similarities are abundantly clear.

We also want to give some feel for the overall topology of the field at the time shown in Figure 2. Rather than simply plotting the components of the fields in the equatorial plane, we also show in Figure 3 a line integral convolution representation [8] of the magnetic field at this time. The line integral convolution method of displaying fields has an advantage over field line or vector field array displays, in that it shows the structure of the field at a resolution close to that of the display. In this display, the "streaks" are parallel or anti-parallel to the direction of the magnetic field. At this time, the magnetic field is only non-zero in the interval $0.5\,a < r < 1.5\,a$. This is consistent with the fact that the information that the spherical surface at $r = a$ has started spinning propagates away from the surface of the sphere at the speed of light, beginning at time $t = 0$, and we are looking at a later time $t = 0.5\,a/c$. We have color-coded Figure 3 so that the red zones correspond to fields associated with the times at which the currents in the sphere are in the process of turning on, and the darker blue zones are the fields associated with the times at which the currents in the spherical shell are constant in time. Note that most of the fields created by the spin-up of the sphere are associated with times after the sphere is rotating at constant angular speed and is no longer accelerating. Figure 2 is one frame of a complete movie that can be found online [6].

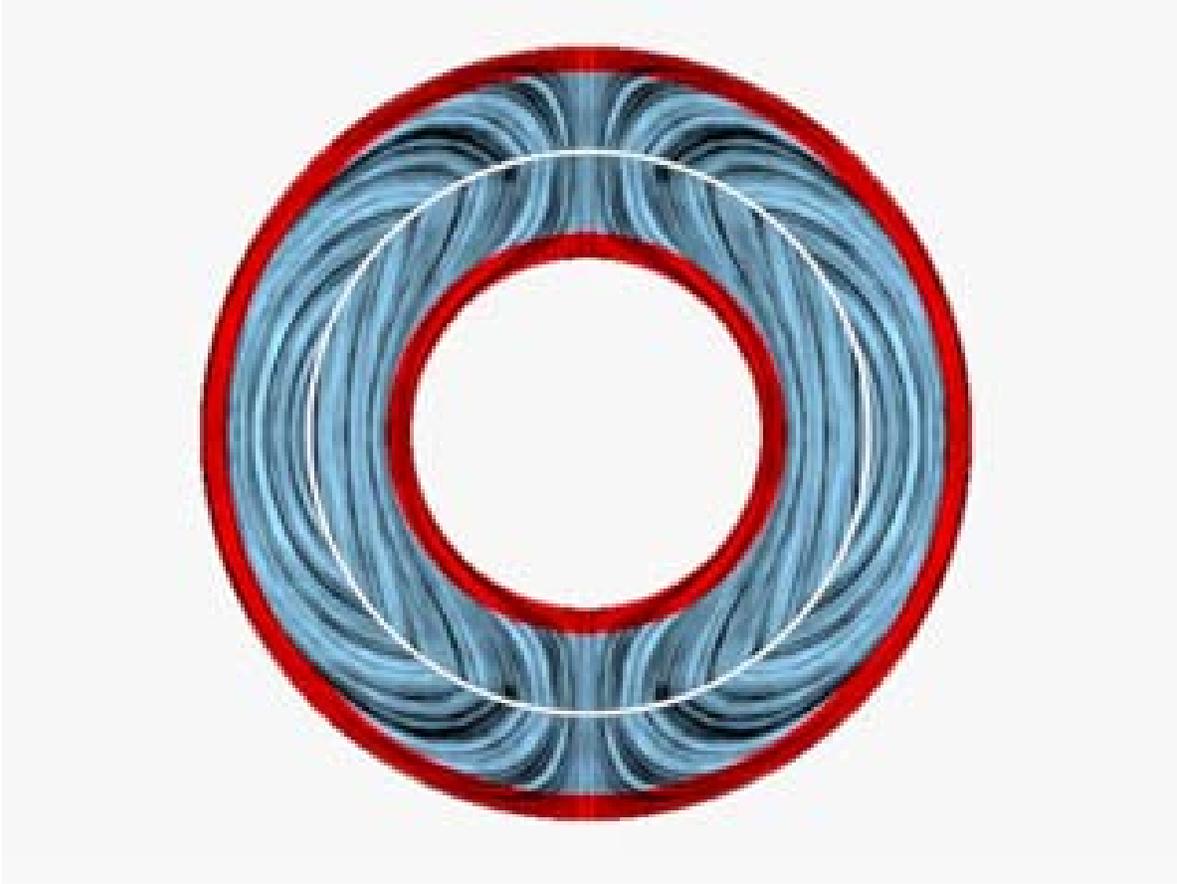

**Figure 3:** A line integral convolution representation of the magnetic field topology for a ramp turn-on in a time $T = 0.1 \ a/c$. We show the field at a time $t = 0.5 \ a/c$ after the turn-on starting at $t = 0$. The sphere is indicated by the silver circle.

To follow this process of an abrupt ramp turn-on to its end, we show in Figure 4 a plot of the same quantities as in Figure 2, for the same fast ramp-up time scale of $T = 0.1 \ a/c$, except now at a much later time $t = 3 \ a/c$. At this time, the electric field of the shell is zero except in the spatial interval $1.9 \ a < r < 4 \ a$. Inside of the distance $r = 1.9 \ a$, we see zero electric field and a static magnetic field which falls off as $1/r^3$ for $a < r < 1.9 \ a$ and for $r < a$ is constant with a value of $B_\theta^{\text{shell}} = -2\mu_o \kappa_o / 3$ in the equatorial plane (corresponding to an upward constant field of $\hat{\mathbf{z}} \, 2\mu_o \kappa_o / 3$). As expected from Ampere's Law, we see a jump in $B_\theta^{\text{shell}}$ across $r = a$ of $\mu_o \kappa_o$, with the sign of the jump consistent with a constant current in the positive $\phi$-direction.

The reason that the electric field is zero for $r < 1.9 \ a$ in Figure 4 is that at $t = 3 \ a/c$, if $r < 1.9 \ a$, the electric field generated by the "far" side of the rotating sphere has had time to propagate across the sphere and exactly cancel out the electric field generated by the "near" side of the rotating sphere. We thus see a "burst" of radiation propagating to infinity with a spatial extent of about $2 \ a$, corresponding to a time duration of about $2a/c$, just as we would expect. At the time shown in Figure 4, the external agents responsible for spinning up the sphere are no longer doing work, since the electric field at the surface of the sphere is zero. These agents only do work in the time interval $0 < t < 2.1a/c$, and in that time they do work sufficient to both establish the static magnetic field and to provide the energy carried off to infinity by the burst of radiation. For $T = 0.1a/c$ these two energies (radiated and stored) are approximately equal, and in the limit that $T$ goes to zero (instantaneous turn-on), they become exactly equal.

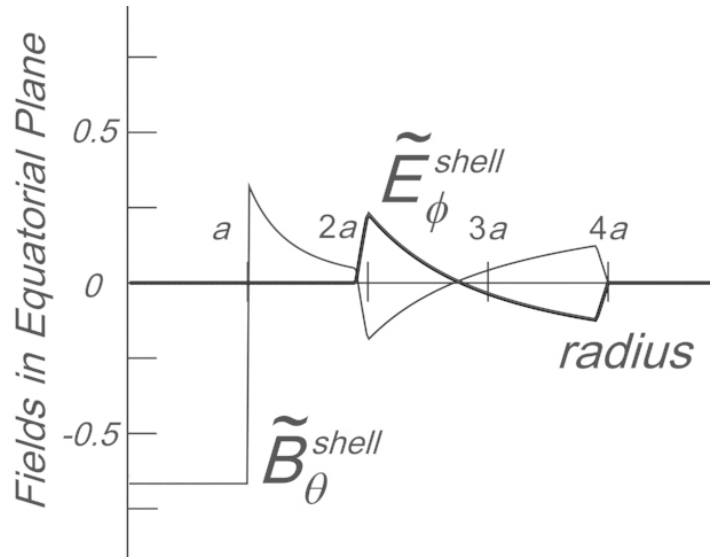

**Figure 4:** The normalized $\phi$ and $\theta$ components of the electric and magnetic field of the shell in the equatorial plane for a ramp turn-on with $T = 0.1\ a/c$, showing the fields at a time $t = 3a/c$ after a turn-on starting at $t = 0$

To give some feel for the overall topology of the field at the time shown in Figure 4, we again show in Figure 5 a line integral convolution representation of the magnetic field at this time. We have color-coded this figure so that the red zones correspond to fields associated with the times when the currents in the sphere are in the process of turning on, the darker blue zones are the fields associated with the times when the currents are constant in time but the agents spinning up the sphere are still doing work, and the light blue zones are the static magnetic field regions, associated with times when the external agents are no longer doing work. At the point in time at which Figure 5 applies, the agents responsible for spinning up the sphere are no longer doing work. The energy required to finish establishing the static magnetic field everywhere in space and to fuel the burst of radiation going to infinity is being transported outward (and locally deposited) by the fields in the red and dark blue regions of Figure 5.

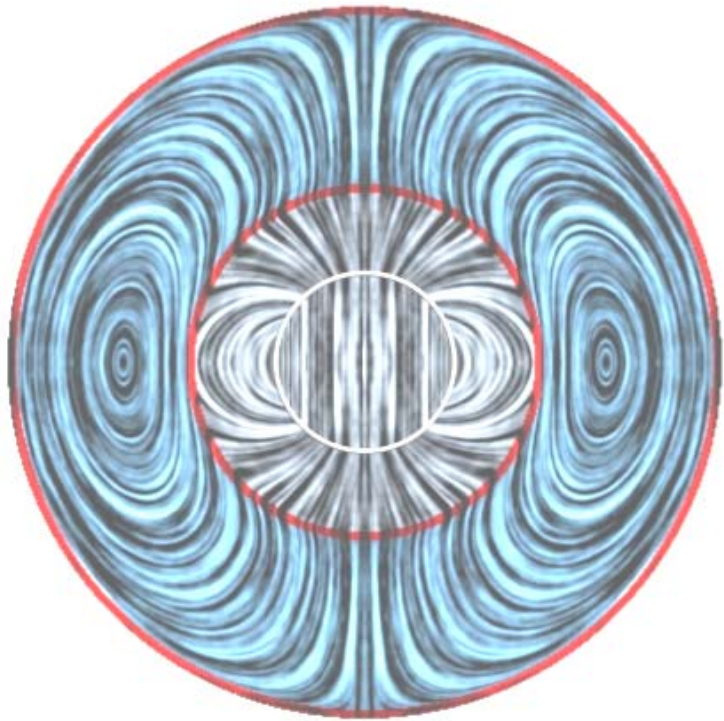

**Figure 5:** A line integral convolution representation of the magnetic field topology for a ramp turn-on in a time $T = 0.1\, a/c$. We show the field at a time $t = 3\, a/c$ after the turn-on starting at $t = 0$.

Finally, consider the total radiated energy versus $a/cT$ for the ramp spin-up of Eq. (41). The function in Eq. (41) is, of course, not expandable in a Taylor series, and the classic expression for the radiated power in the dipole limit, proportional to the square of the second time derivative of the magnetic dipole moment, cannot be evaluated. However there is no problem in evaluating our complete spherical shell expressions using Eq. (41) to compute the fields, because we can easily integrate the ramp successive times unambiguously. We show that numerical calculation for the total energy radiated in the ramp spin-up in Figure 1, in the curve labeled "Ramp". The behavior at low values of $a/cT$ is now proportional to $(a/cT)^2$, and this is understandable analytically (with some effort). The behavior at high values of $a/cT$ is the same as for the "Smooth" function, approaching the same constant, as we expect, since in this limit both $m_{ramp}(t)$ and $m_{smooth}(t)$ approach a step function at $t = 0$.

## 5 Summary

Studies of the electromagnetic fields everywhere in space for current distributions with arbitrary time dependence and at least one finite and non-zero spatial dimension are difficult. Pedagogically, however, even in intermediate level courses, we would like a student to be able to observe an abrupt change in current in one part of such a system causing changes in the local fields, and to observe these field changes as they propagate at the speed of light to other parts of the same systems of currents, and interact with those currents.

In this paper, we have given the complete solutions everywhere in space for two such systems, a planar solenoid and a spherical shell of current. Unexpectedly, we find that our solutions for these electromagnetic fields are free of the concepts of differential calculus, in that our solutions only involve the currents and their time integrals, and do not involve the time derivatives of the currents. These problems are of physical interest because: (1) they show that current systems of finite extent can radiate even during time intervals when the currents are constant; (2) they explicitly display transit time delays across a source associated with its finite dimensions; and (3) they allow students to see directly the origin of the reaction forces for time-varying systems; (4) they allow for the direct calculation of the ratio of the energy radiated to the energy stored for arbitrary time dependence.

We present four problems below which bear on this understanding, three of which are appropriate for the undergraduate level, and the fourth of which is appropriate for a graduate level course in classic electromagnetism.

# 6    Sample Problems

## 6.1    The planar solenoid and Faraday's Law

This problem is appropriate for an intermediate level undergraduate course, that is, the typical junior/senior level course taken by physics majors. The problem probes the quasi-static approximation for the magnetic field of the planar solenoid. In this approximation, we ignore propagation effects and consider the magnetic field inside the solenoid to be spatially uniform and the magnetic field outside the solenoid to be zero. Since we have the exact solutions for the planar solenoid, we can compare those exact solutions to the fields in the quasi-static approximation and draw conclusions about the validity of the approximation. In particular, we will compare our exact solutions to the calculation of the electric field at the surface of the solenoid using Faraday's Law applied assuming the quasi-static approximation for the magnetic field.

We have two current sheets in the $yz$ plane as described above. We assume that in a previous problem the student has been led through the derivation of Eq. (9) above.

(a)  Assume that to a good approximation the magnetic field as a function of time and space is given by $\mathbf{B}(x,t) = \mu_o \kappa(t)\hat{\mathbf{z}}$ for $|x| \le a$ and zero otherwise. With this assumption, use Faraday's Law to find the electric field at $x = \pm a$.

(b)  Find an expression for $\mathbf{E}(x,t)$ from the exact solution for the vector potential in Eq. (9). Assuming that the time scale $T$ for significant variation in $\kappa(t)$ is much greater than $a/c$, expand your exact expression for $\mathbf{E}$ using Taylor series, keeping only leading order terms in the small quantity $a/cT$. Show that if you keep terms to this order, then the exact solution for the electric field assuming $T >> a/c$ reduces to the electric field you obtained in (a) at $x = \pm a$.

## 6.2 The planar solenoid problem comparing radiated and stored energy

We have two current sheets in the $yz$ plane as described above. We assume that in a previous problem the student has been led through the derivation of Eq. (9) and the associated fields.

(a) Assuming that $T >> a/c$, show that for $x > a$ Eq. (9) can be written to leading order as

$$\mathbf{A}(x,t) = \hat{\mathbf{y}}\, c\mu_o a\, \kappa\left(t - \frac{x}{c}\right) \tag{42}$$

(b) Assume the functional form for $\kappa(t)$ is similar to the "smooth" turn-on form given in Eq.(35), and that $T >> a/c$. Using your solution for the vector potential in Eq. (42), and its associated fields, compute the ratio of the energy radiated away to infinity per unit $yz$ area of the solenoid to the energy in the final magnetic field per unit $yz$ area of the solenoid for $t > T$ after the current has stopped increasing in time (Hint: $\int_{-\infty}^{\infty} \frac{d\eta}{\left(1+\eta^2\right)^2} = \frac{\pi}{2}$). [Answer: $\frac{5}{\pi} a/cT$].

(c) Now assume that $\kappa(t)$ is a step function, that is, $\kappa(t)$ goes instantaneously from 0 to $\kappa_o$ at $t = 0$. Use the fields derived from your exact solution for the potential in Eq. (9) to compute the same ratio as in (b) above. [Answer: 1]

## 6.3 Reaction forces for the uniformly-charged rotating spherical shell

We have a spherical shell carrying a current as described in Eq.(10). We assume that in a previous problem the student has been led through the derivation of Eq. (19) using the time domain method (see the development in [6]).

(a) In the case that the time scale $T$ for significant variation in $m(t)$ is much greater than $a/c$, use Taylor series expansions of the exact solutions in Eq. (19) to show that the electric field at $r = a$ is given by Eq. (33).

(b) Assume that current is turned on over a time $T$ with $\kappa(t) = 0$ for $t < 0$ and $\kappa(t) = \kappa_o$ for $t > T$. Assume also that the turn on and the leveling off to constant current are smooth in the sense that the first through third time derivatives of $\kappa(t)$ vanish both at $t = 0$ and at $t = T$. Using Eq. (10) and Eq. (33), calculate the work done (that is, the time and space integral of $-\mathbf{J} \cdot \mathbf{E}$) in the case that the current varies slowly in the sense that the time-scale for change $T$ is much larger than $a/c$. Relate your answer to the energy stored in the magnetic field after $t = T$ (see Eq. (36)) and to the energy radiated away in magnetic dipole radiation between $0$ and $T$. [Hints: the rate at which energy is radiated into all solid angles in magnetic dipole radiation is given by $\mu_o \ddot{m}^2 / 6\pi c^3$, and the magnetic dipole moment for this problem is given in Eq.(11)]

(c) Now assume that $\kappa(t)$ goes instantaneously from 0 to $\kappa_o$ at $t = 0$. Consider only the radiation terms in the electric field associated with the potential in Eq.(19), that is the terms going as $1/r$, and for these terms assume that the corresponding $B$ field radiation terms are in magnitude just the $E$ field radiation terms divided by $c$. Using your exact solutions, compute the energy radiated away to infinity for this process of instantaneously spinning up the sphere, and show that it is equal to the energy stored in the magnetic field long after the current has stopped increasing in time.

## 6.4   The spherical shell graduate level problem

Consider a current carrying shell in which the current in the $\phi$-direction depends on the polar angle $\theta$ in the following manner.

$$\mathbf{J}(r,\theta,t) = \hat{\boldsymbol{\phi}}\kappa(t)\delta(r-a)P_l^1(\cos\theta) \tag{43}$$

where $P_l^m(\cos\theta)$ is an associated Legendre polynomial.

(a) Show that the solution for the vector potential $\mathbf{A}$ for this problem is given by

$$A_\phi(r,t) = \frac{3c\mu_o}{8\pi a^3 r}P_l^1(\cos\theta)\sum_{k=0}^{l}\sum_{m=0}^{l}\frac{\gamma_{l,k}\gamma_{l,m}c^{k+m}}{(2r_>)^k(2r_<)^m}\begin{bmatrix}(-1)^m m^{(k+m+1)}(t_+)\\+(-1)^{l+1}m^{(k+m+1)}(t_-)\end{bmatrix} \tag{44}$$

where $m^{(n)}(t)$ is defined in Eq. (20) and Eq. (21), and

$$\gamma_{l,m} = \frac{(l+m)!}{m!(l-m)!} \tag{45}$$

(b) How would you solve the general problem for any azimuthally symmetric current distribution on a spherical shell by superposition of the solutions given in Eq.(44)?

(c) Now assume that $\kappa(t)$ goes instantaneously from 0 to $\kappa_o$ at $t = 0$. Consider only the radiation terms in Eq.(44). Using your exact solutions, show that the energy radiated away to infinity for this process of instantaneously spinning up the sphere is equal to the energy stored in the static magnetic field a long time after the current has stopped increasing in time. That energy in the static magnetic field for general $l$ is given by

$$U_l^{mag} = \frac{l(l+1)}{(2l+1)^2} \frac{9\mu_0 m_o^2}{8\pi a^3} \tag{46}$$

(d)  Argue from energy considerations that for any azimuthally symmetric current distribution on a spherical shell, the energy required to establish the currents in a time very short compared to $a/c$ must be approximately equal to twice the energy stored in the magnetic field after they the currents are established.

(e)  Can you think of a way to show (d) directly from your solution in Eq. (44), rather than relying on an energy argument?

## 7    Acknowledgments

RHP acknowledges support from NSF grant PHY0554367.

## 8    References

[1] Jackson, J. D., *Classical Electrodynamics*, 2<sup>nd</sup> Edition (John Wiley and Sons, 1975) see Chapter 16.

[2]  H. Spohn, *Dynamics of charged particles and their radiation field* (Cambridge University Press 2004); W. Appel and M. Kiessling, "Scattering and Radiation Damping in Gyroscopic Lorentz Electrodynamics," Lett. Math. Phys. **60**, 31-46 (2002) and "Mass and Spin Renormalization in Lorentz Electrodynamics," Annals Phys. **289**, 24-83 (2001); M. Kunze, "On the absence of radiationless motion for a rotating classical charge" Adv. Math., **223**, 1632-1665 (2010).

[3]  R. P. Feynman, R. B. Leighton, and M. Sands, *The Feynman Lectures on Physics*, Vol. II, Sec. 18-4 (Addison-Wesley, Reading, MA, 1964).  See also: J.-M. Chung, "Revisiting the radiation from a suddenly moving sheet of charge,"  Am. J. Phys. **76**(2), 133-136 (2008); Barry R. Holstein, "Radiation from a suddenly moving sheet of charge," Am. J. Phys. **63**(3), 217–221 (1995);  P. C. Peters, "Electromagnetic radiation from a kicked sheet of charge," Am. J. Phys. **54**(3), 239– 245 (1986), see in particular Eq. (6) for the electric field of a single sheet of moving charge and the accompanying discussion in Sec. II of this paper; T. A. Abbott and D. J. Griffiths, "Acceleration without radiation," Am. J. Phys. **53**(12), 1203-1211 (1985), see in particular, Section III of this paper.

[4] An $x$ component of the electric field would arise from the scalar potential $\varphi$ and from the relationship $\mathbf{E} = -\nabla\varphi - \partial\mathbf{A}/\partial t$ .  We ignore $\varphi$ and hence omit the $x$-component of $\mathbf{E}$. Since we treat the charge density as a constant in time the scalar potential and $E_x$ would be constant, and hence irrelevant to our considerations.  Note that we could have chosen to have the surface  current consist of opposite charge densities driven into motion in opposite directions, producing the same total current density as in Eq. (2), but without any charge density, and hence without any $E_x$.  These same considerations, with

appropriate modifications justify our omission of the radial electric field for the spinning spherical shell in Sec. 3.

[5]  Daboul, J., and J. D. Jensen, "Radiation reaction for a rotating sphere with rigid surface charge,"  Z. Physik. **265**, 455-478 (1973), see particularly Eq. (2.23) and (2.24) of this paper, and Vlasov, A. A., "Radiation reaction in classical electrodynamics: the case of a rotating charged sphere," arXiv:physics/9801017v1[physics.class-ph] (1998), see particularly Eq. (8) of this paper.

[6]  The methods for solving the spherical case are discussed in a technical note accompanying this paper.  This technical note also contains the solutions to the problems posed in the text, and is available on the web at

http://web.mit.edu/viz/spin/visualizations/theTheory/theTheory.htm

Movies of Figures 2 through 5, may be found at  http://web.mit.edu/viz/finalDrafts/. Other movies can be found at http://web.mit.edu/viz/spin/.   A preprint relevant to the material considered here is "The electromagnetic fields of a spinning shell of charge" by S. Olbert and J. W. Belcher, see  http://arxiv.org/abs/1010.1917.

[7] Reference [1], see Eq. (9.33), which gives the Fourier domain solution for the vector potential.  With the replacement $ik = i\omega / c$  by $-\left(1/c\right)d / dt$, we recover our Eq. (27).

[8]  B. Cabral and C. Leedom, "Imaging vector fields using line integral  Proc. SIGGRAPH '93, 263-270 (1993).